\documentclass{ws-ijmpcs}

\begin{document}

\markboth{T. Ledwig}
{Instructions for Typing Manuscripts (Paper's Title)}

%
\catchline{}{}{}{}{}
%

\title{The nucleon mass and pion-nucleon sigma term from a chiral analysis of $N_f = 2+1$ lattice QCD world data}

\author{L. Alvarez-Ruso, T. Ledwig, M.J. Vicente Vacas}
\address{Departamento de F\'\i sica Te\'orica and IFIC, Centro Mixto
  Universidad de Valencia-CSIC,\\Valencia, Spain\\ledwig@ific.uv.es}


\author{J. Martin-Camalich}
\address{Department of Physics and Astronomy, University of Sussex,\\Brighton, UK}


\maketitle


\begin{abstract}
  Fits of the $p^4$ covariant $SU(2)$ baryon chiral perturbation theory 
  to lattice QCD nucleon mass data from several collaborations for 2 and 2+1 flavors are presented. 
  We consider contributions from explicit $\Delta\left(1232\right)$ 
  degrees of freedom, finite volume and finite spacing corrections.
  We emphasize here on our $N_f=2+1$ study.
  We obtain low-energy constants of natural size that are compatible with the rather 
  linear pion-mass dependence of the nucleon mass observed in lattice QCD. 
  We report a value of $\sigma_{\pi N}=41(5)(4)$~MeV  in the 2 flavor
  case and $\sigma_{\pi N}=52(3)(8)$ MeV for 2+1 flavors.
\end{abstract}

\section{Introduction}	
Through lattice QCD simulations (lQCD) it is possible to study QCD at current quark masses which are not restricted to 
their physical values. Data points obtained for such a scenario represent a powerful 
pool of information to fix free low-energy constants (LECs) of the (baryon) chiral perturbation theory (B$\chi$PT); 
LECs that also enter calculations for experimental observables. 
Thus, adjusting the B$\chi$PT to lQCD data in the unphysical region allows for predictions at the physical point.

In our study we performed such a matching for the quark-mass ($m_{u}=m_{d}=\overline{m}$) dependence of the nucleon mass 
$M_N(\overline{m})$. We fit lQCD data for the $N_f=2$ flavors, two light degenerated quarks, as well as
for the $N_f = 2+1$ ensembles, two light degenerated and one heavy quarks. Here, we concentrate 
on our $N_f=2+1$ results and refer to Ref. \cite{paper} for more details on the $N_f=2$ ones.

A measure of the contribution from the explicit chiral symmetry breaking to the nucleon mass is given by 
the so-called $\sigma_{\pi N}$-term. It is the nucleon scalar form factor at zero momentum transfer squared which 
can be isolated from $\pi N$-scattering experiments. Alternatively, the $\sigma_{\pi N}$ can 
also be obtained by the Hellmann-Feynman theorem from $M_N(\overline{m})$:
\begin{equation}
  \overline{m}\frac{\partial}{\partial\overline{m}}M_{N}\left(\overline{m}\right)=\sigma_{\pi N}=
  \overline{m}\langle N |\overline{u}u+\overline{d}d| N \rangle\,\,\,\,.
\label{eq:HF}
\end{equation}
As a result, the function $M_N(\overline{m})$, which has mainly unphysical values,
connects lQCD, B$\chi$PT and experiment.

\section{Nucleon mass and covariant baryon chiral perturbation theory}
The B$\chi$PT up to $p^4$ describes the nucleon mass by:
\begin{equation}
  M_{N}^{\left(4\right)}\left(M_{\pi}^{2}\right)  =  M_{0}-c_{1}4M_{\pi}^{2}+
  \overline{\alpha}\frac{M_{\pi}^{4}}{2}+\frac{c_{1} M_\pi^4}{8\pi^{2}f_{\pi}^{2}}
  \ln\frac{M_{\pi}^{2}}{M_{0}^{2}}+\Sigma_{loops}^{\left(3+4\right)}\left(M_{\pi}^{2}\right)
  +\mathcal{O}\left(p^{5}\right),\label{eq:MN(4)(Mpi)}
\end{equation}
with $M_\pi^2 \sim \overline{m}$ and $f_\pi$ as the pion mass and pion decay constant. We fit the 
three LECs $c_1$, $\overline{\alpha}$ and $M_0$ to lQCD data 
and obtain through Eq. (\ref{eq:HF}) a $\sigma_{\pi N}$ value. Explicit
$\Delta(1232)$ contributions can appear in the loop-contributions 
$\Sigma_{loops}^{\left(3+4\right)}$ and we refer to Ref. \cite{paper} for all technical details. 
Generally, lQCD data is given in the dimensionless form $(aM_\pi, aM_N)$ where each collaboration 
sets its scale through a specific value for the lattice spacing $a$.
In the $N_f=2+1$ case, we fit the data after converting to physical units. Since all 
three quantities are subject to statistical uncertainties, the $a$ uncertainty translates into a 
correlated uncertainty for the normalization of data points belonging to the same set. 
We take this correlation into account by defining a correlation matrix $V$ containing all 
these uncertainties, and minimize the function:
\begin{equation}
\label{eq:chi1}
  \chi^{2}  =  \vec{\Delta}^{T}V^{-1}\vec{\Delta}\,\,\,\mbox{,}\,\,\,\Delta_{i}=M_{N}^{(4)}\left(M_{\pi,i}^{2}\right)+c_{i}a_{i}^{2}+\Sigma_{N}^{(4)}\left(M_{\pi,i}^{2},L_{i}\right)-d_{i}\left(M_{\pi,i}^{2},L_{i}\right),
\end{equation}
with $d_i$ being the data and $M_N^{(4)}$ and $\Sigma^{(4)}$ the B$\chi$PT infinite volume nucleon mass and
its finite volume corrections. The constants $c_i$ parametrize finite spacings effects for each 
lQCD-action separately
In our $SU(2)$ B$\chi$PT analysis we only include $N_f=2+1$ lQCD data for which the strange quark mass
is kept constant approximately at its phenomenological value. For this case the strange quark mass corrections
to the nucleon mass can be seen as being integrated out into the LECs we are fitting.
Furthermore, to ensure controllable finite volume corrections and a reasonable chiral convergence of our 
results, we only account for data points with $M_\pi L>3.8$ and $M_\pi<415$ MeV, with $L$ being the lattice box size.
Following this, we include data from the BMW \cite{BMW}, HSC \cite{HSC}, LHPC \cite{LHPC}, 
MILC \cite{MILC}, NPLQCD \cite{NPLQCD}, PACS \cite{PACS} and RBCUK-QCD \cite{RBCUK-QCD} collaborations.

\section{Results and conclusion}
\begin{figure}[h]
\centering
\includegraphics[width=6.5cm,clip]{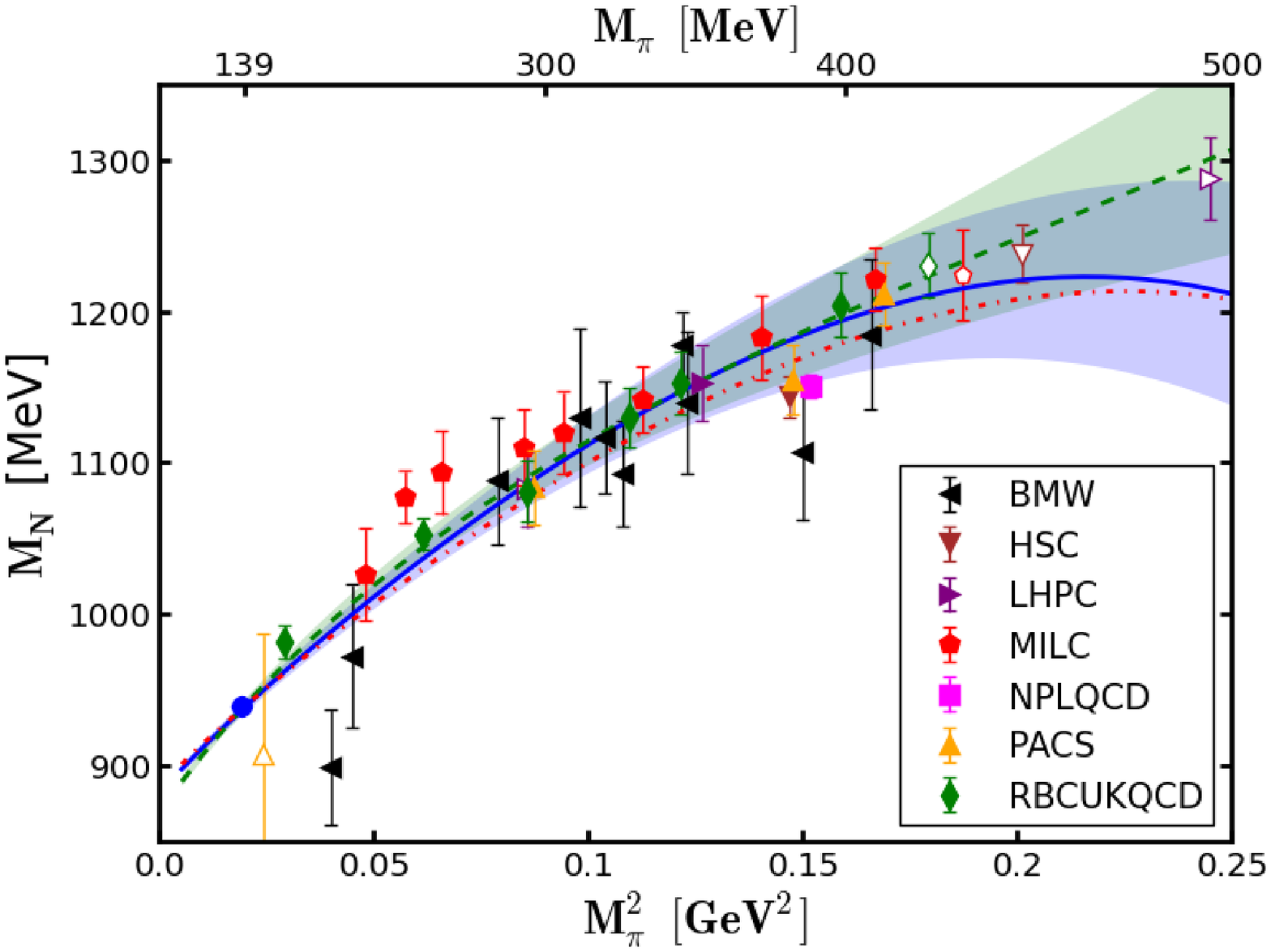}\includegraphics[width=6.5cm,clip]{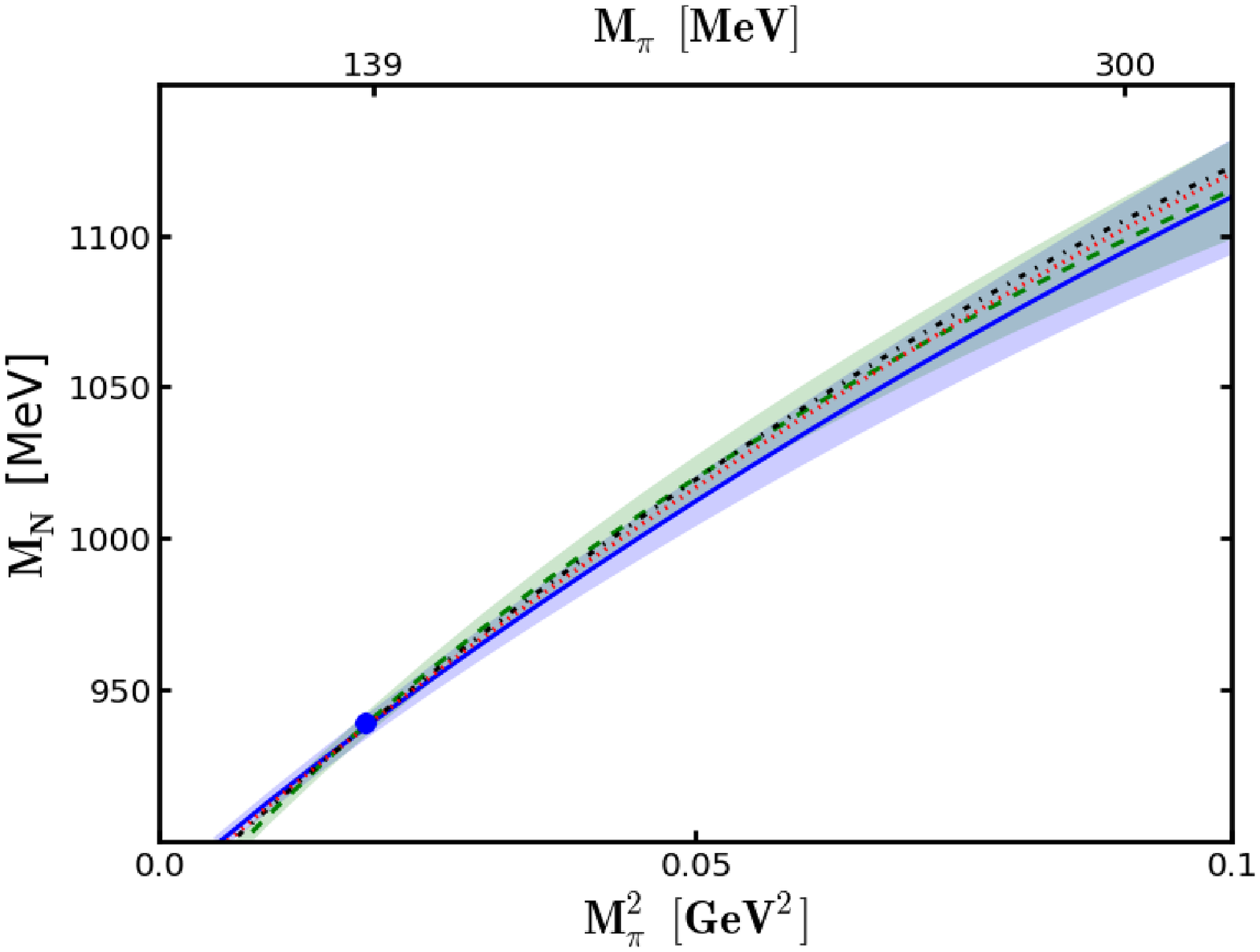}
\caption{Pion-mass dependence of the nucleon mass from $N_f=2+1$ fits. Right: 
The blue-solid (green-dashed) line shows our
result for the $\mathcal{O}(p^4)$ B$\chi$PT fit with(out) $\Delta(1232)$ contributions by excluding the two data 
points from the HSC and NPLQCD collaborations. The red-dotted line is a B$\chi$PT fit with
 $\Delta(1232)$ including also this data. Left: Same results depicted for a smaller
$M_\pi$ range.}
\label{fig-1}     
\end{figure}
\begin{figure}[h]
\centering
\includegraphics[width=6.5cm,clip]{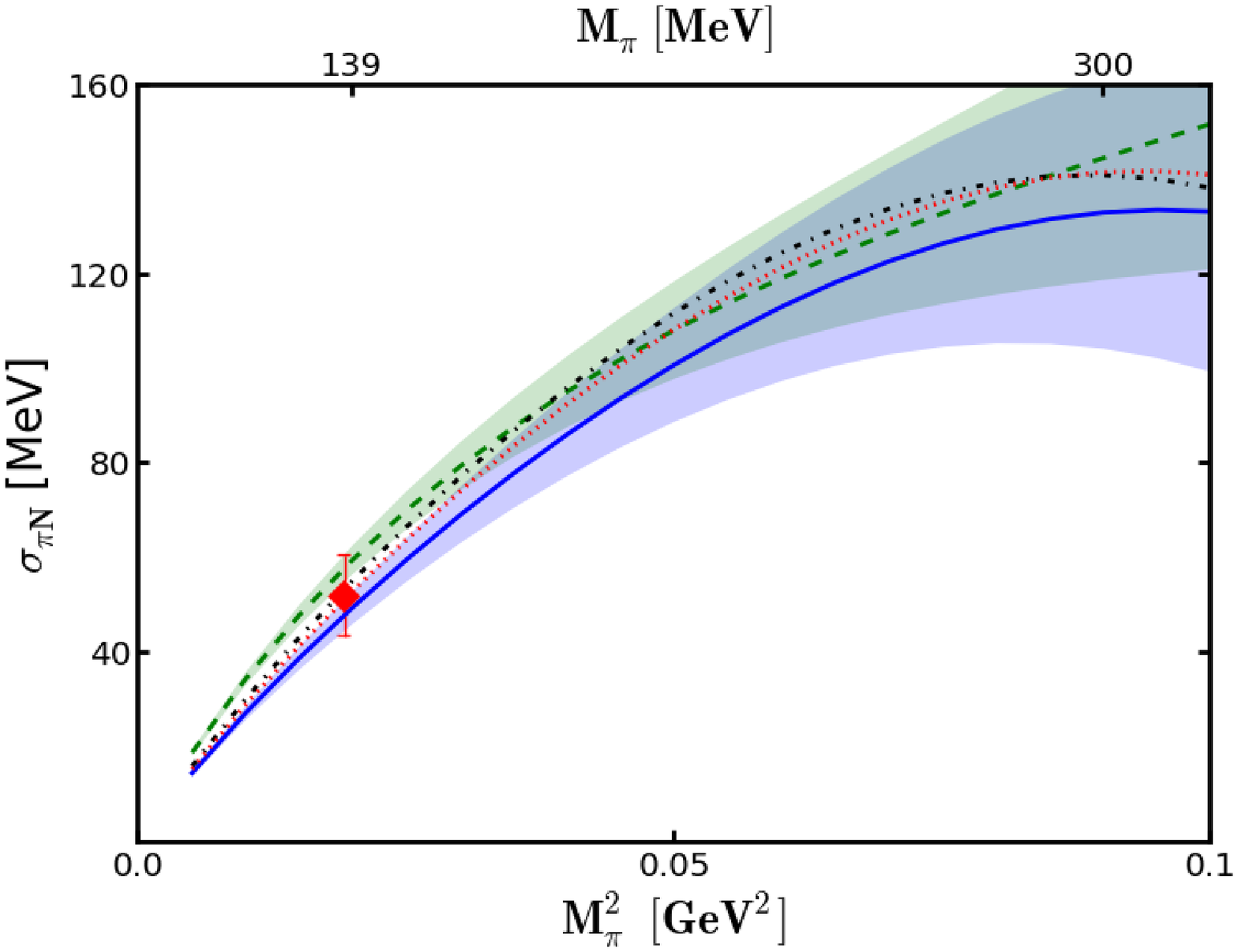}\includegraphics[width=6.5cm,clip]{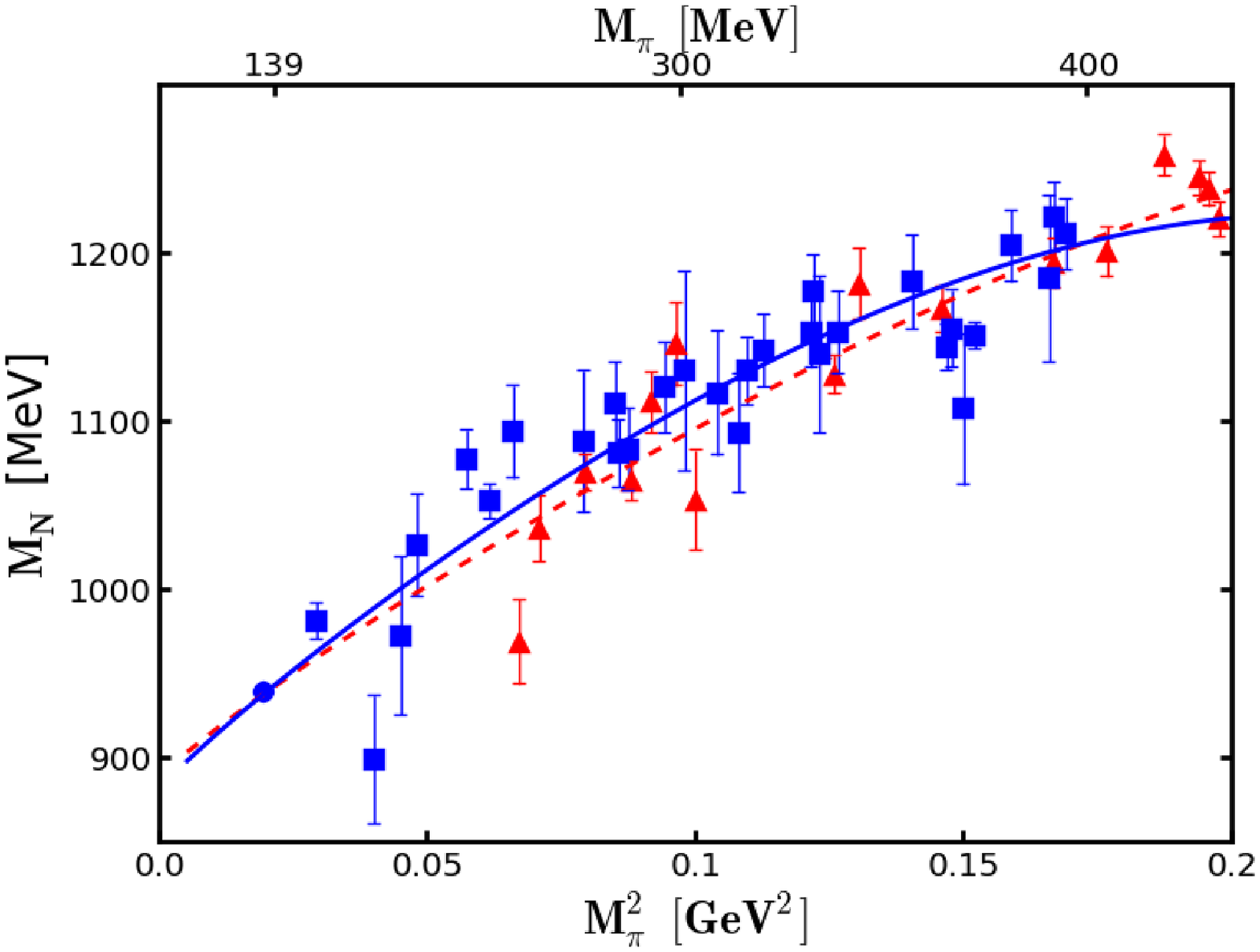}
\caption{Left: Pion-mass dependence of the $\sigma_{\pi N}$ term from $N_f=2+1$ fits. The color
  code is the same as in Fig. \ref{fig-1}. The red-diamond is our predicted $\sigma_{\pi N}$ value 
  at the physical point. Right: Comparison of B$\chi$PT $p^4$ fits with $\Delta(1232)$ to $N_f=2$ 
  (red-triangles) and $N_f=2+1$ (blue-squares) lQCD data. The blue-circle is the physical nucleon mass.}
\label{fig-2}     
\end{figure}

The Fig. \ref{fig-1} shows several B$\chi$PT fit strategies to the above listed $N_f=2+1$ lQCD data. 
We obtain consistent descriptions of the pion-mass dependence of the nucleon mass 
where the slope-variations are below the resolution of the current data.
However, by looking to the pion-mass dependence of the $\sigma_{\pi N}$ term, left graph of Fig. \ref{fig-2},
we observe that these small variations translate into noticeable changes.
For successively including explicit $\Delta(1232)$ contributions and finite 
spacing corrections in our fit formula, the $\sigma_{\pi N}$ term at the physical point changes by up 
to $14$ MeV from $\sigma_{\pi N} = 58(3)$ MeV to $49(2)$MeV down to $44(3)$ MeV.
In view of this, we take a conservative standpoint and give the following weighted average
as our final $N_f=2+1$ $\sigma_{\pi N}$ value at the physical point:
\begin{equation}
\sigma_{\pi N}^{N_f=2+1}=52\left(3\right)\left(8\right)\,\,\,\,\mbox{MeV}\,\,,
\end{equation}
where the second (systematical) uncertainty spans all the above central values.

The right graph of Fig. \ref{fig-2} compares our B$\chi$PT results from separate fits
to lQCD data from $N_f=2$ and $N_f=2+1$ ensembles. Also here, differences are 
within the spread and uncertainties of the input data but result
in a slightly smaller $\sigma_{\pi N}$ value for the $N_f = 2 $ case:
\begin{equation}
\sigma_{\pi N}^{N_f=2}=41\left(5\right)\left(4\right)\,\,\,\,\mbox{MeV}\,\,.
\end{equation}
Concerning the uncertainties, these two values are compatible, however, show a slight tension which we 
mostly trace back to the different data point distributions.
On the one hand, the current $N_f=2$ data does not constrain the low-$M_\pi$ region as much as the 
$N_f=2+1$ data. This region is dominated by the LEC $c_1$ which is the $\sigma_{\pi N}$ at
leading order. Consequently, we obtain by $\sim 13$ \% different $c_1$-values with correspondingly 
different contributions to the total $\sigma_{\pi N}$. On the other hand, in the $N_f=2$ case one direct 
data point \cite{QCDSF} of the $\sigma_{\pi N}$ at $M_\pi \approx 290$ MeV could be included in the fit. 
A positive outcome was that the above spread of our results with and without inclusion of $\Delta(1232)$ 
contributions is lifted. Since no such data point at low-$M_\pi$ is available for the $N_f=2+1$ case, 
a spread of $9$ MeV remains. Note that in the HB$\chi$PT formalism \cite{HBChPT} this spread is more than $40$ MeV.
To compare our above numbers also with phenomenology, 
we cite the latest value extracted from pure $\pi$-N scattering data: $\sigma_{\pi N} = 59(7)$ MeV \cite{piN}.
In conclusion, we performed fits of the $p^4$ $SU(2)$ covariant B$\chi$PT with and without
explicit $\Delta(1232)$ contributions to lQCD nucleon mass data. Even though the present data set is
extensive, we spotted systematic effects coming mainly from the distribution of the data points.
These systematic effects would further be reduced by new data for: a) more $N_f=2$ data points in the low-$M_\pi$ region, 
b) already one direct calculation of the $\sigma_{\pi N}$ at $M_\pi<300$ MeV for the $N_f = 2 + 1$ case.
\section*{Acknowledgments}
The work has been supported by the Spanish Ministerio de Economía y Competitividad and 
European FEDER funds under Contracts FIS2011-28853-C02-01 and FIS2011-28853-C02-02, Generalitat Valenciana 
under contract PROMETEO/2009/0090 and the EU Hadron-Physics3 project, Grant No. 283286. JMC also 
acknowledges support from the Science Technology and Facilities Council (STFC) under grant ST/J000477/1,
the grants FPA2010-17806 and Fundaci\'on S\'eneca 11871/PI/09.

\end{document}